\documentclass[reprint,aip,apl]{revtex4-1}
\usepackage{graphicx}
\usepackage{dcolumn}
\usepackage{bm}
\usepackage{subfigure}

\begin{document}

\title{Electrocaloric effect in BaTiO$_3$: a first-principles-based
  study on the effect of misfit strain}

\author{Madhura Marathe} 
\email{madhura.marathe@mat.ethz.ch}
\author{Claude Ederer} 
\email{claude.ederer@mat.ethz.ch}
\affiliation{Materials Theory, ETH Z\"urich, Wolfgang-Pauli-Str. 27,
  8093 Z\"urich, Switzerland} 
\date{\today}

\begin{abstract}
We address the question of how the electrocaloric effect in epitaxial
thin films of the prototypical ferroelectric BaTiO$_3$ is affected by
the clamping to the substrate and by substrate-induced misfit
strain. We use molecular dynamics simulations and a
first-principles-based effective Hamiltonian to calculate the
adiabatic temperature change $\Delta T$ under different epitaxial
constraints. Our results demonstrate that, consistent with
phenomenological theory, clamping by the substrate reduces the maximum
$\Delta T$ compared to bulk BaTiO$_3$. On the other hand, compressive
misfit-strain leads to a strong increase of $\Delta T$ and shifts the
maximum of the electrocaloric effect to higher temperatures. A rather
small compressive strain of $-0.75$\,\% is sufficient to obtain a
$\Delta T$ that is larger than the corresponding bulk value.
\end{abstract}

\maketitle

The electrocaloric (EC) effect manifests itself as temperature change
of a dielectric material induced by an applied electric
field.\cite{Scott_2011,Valant_2012} The electric field increases the
order of the electric dipoles, which decreases the entropy of the
system and under adiabatic conditions leads to an increase in
temperature. When the field is removed the dipoles disorder, which,
depending on the process conditions, leads to an increase in entropy
and/or a reduction in temperature.

Even though the EC effect has been known for a long time (see
e.g. Ref.~\onlinecite{Scott_2011}), the observed temperature changes
were considered too small to be suitable for applications. However,
the recent report of a ``giant electrocaloric effect'' in
Pb(Zr,Ti)O$_3$ thin films by Mischenko \textit{et
  al}.~\cite{Mischenko_et_al_2006} has stimulated extensive work in
this area, and has established the EC effect as an attractive
alternative for the design of future cooling
devices.\cite{Faehler_et_al:2011}

The large EC temperature change that is observed in thin films is
mostly due to the larger electric fields that can be applied compared
to bulk ceramics.~\cite{Valant_2012} However, it is also well known
that the ferroelectric properties of thin film materials are strongly
affected by substrate-induced clamping and misfit
strain.\cite{Dawber/Rabe/Scott:2005,Schlom_et_al:2007} For example,
thin films of BaTiO$_3$ (BTO), a typical textbook model ferroelectric,
exhibit strongly enhanced ferroelectric
properties.~\cite{Choi_et_al_2004} Furthermore, epitaxial strain not
only affects the electric polarization and ferroelectric transition
temperatures, but also results in a different sequence of phase
transitions compared to those observed in the unstrained bulk
case.~\cite{Pertsev_Zembilgotov_Tagantsev_1999,Dieguez_et_al_2004}
Using phenomenological Landau-Devonshire theory, it has been shown
that epitaxial strain also affects the EC temperature change in BTO
and related
materials.~\cite{Akcay_et_al_2008,Qiu_Jiang_2008,Qiu_Jiang_2009,Zhang_et_al_2011}

Here, we use a first principles-based effective
Hamiltonian~\cite{Zhong_Vanderbilt_Rabe_1994,Zhong_Vanderbilt_Rabe_1995,Nishimatsu_et_al_2008}
to study the effect of substrate-induced clamping and compressive
epitaxial strain on the EC temperature change in BaTiO$_3$. In the
effective Hamiltonian, the structural degrees of freedom are described
via a small number of collective modes: a polar \emph{soft mode}
related to the ferroelectric distortion in each unit cell, and several
local and global strain variables. All parameters of the corresponding
energy expression can be calculated from first principles electronic
structure calculations. This allows to perform temperature-dependent
simulations for relatively large system sizes without having to resort
to empirical fitting of experimental data.
The effective Hamiltonian approach has previously been used to study
the EC effect in several bulk
materials,~\cite{Prosandeev_Ponomareva_Bellaiche_2008,Lisenkov/Ponomareva:2009,Ponomareva/Lisenkov:2012,Beckman_et_al_2012}
and good agreement with experimental observations has been found.

We perform molecular dynamics (MD) simulations for the effective
Hamiltonian using the FeRAM code developed by Nishimatsu, \textit{et
  al}.~\cite{Nishimatsu_et_al_2008} To simulate BTO, we use the set of
parameters obtained in Ref.~\onlinecite{Nishimatsu_et_al_2010} using
the generalized gradient approximation (GGA) according to Wu and
Cohen.\cite{Wu/Cohen:2006} It has been shown that this
exchange-correlation energy functional leads to more accurate
structural parameters for BTO than previous calculations based on the
local density approximation (LDA).\cite{Nishimatsu_et_al_2010} We
note, however, that in contrast to
Ref.~\onlinecite{Nishimatsu_et_al_2008} we do not apply any empirical
pressure corrections in our calculations.

In order to simulate clamping to a (001)-oriented subtrate, we fix
certain components of the homogeneous strain tensor in the effective
Hamiltonian ($\eta_1 = \eta_2 = \eta = \text{const.}$ and $\eta_6 = 0$
in standard Voigt notation), while all other components as well as all
inhomogeneous strain variables are optimized with respect to the
current soft mode configuration in each MD step. We use a simulation
box of 16$\times$16$\times$16 perovskite unit cells and periodic
boundary conditions in all three cartesian directions, i.e. our
simulations correspond to ``strained bulk''. This allows us to focus
on the pure strain effect while excluding any finite size, surface, or
interface effects. We use 60,000 MD steps to thermalize the system and
then a further 40,000 steps to average over the required properties
with a time step of 2\,fs per MD iteration.

\begin{figure}[tb]
\centering
\includegraphics[width=\columnwidth]{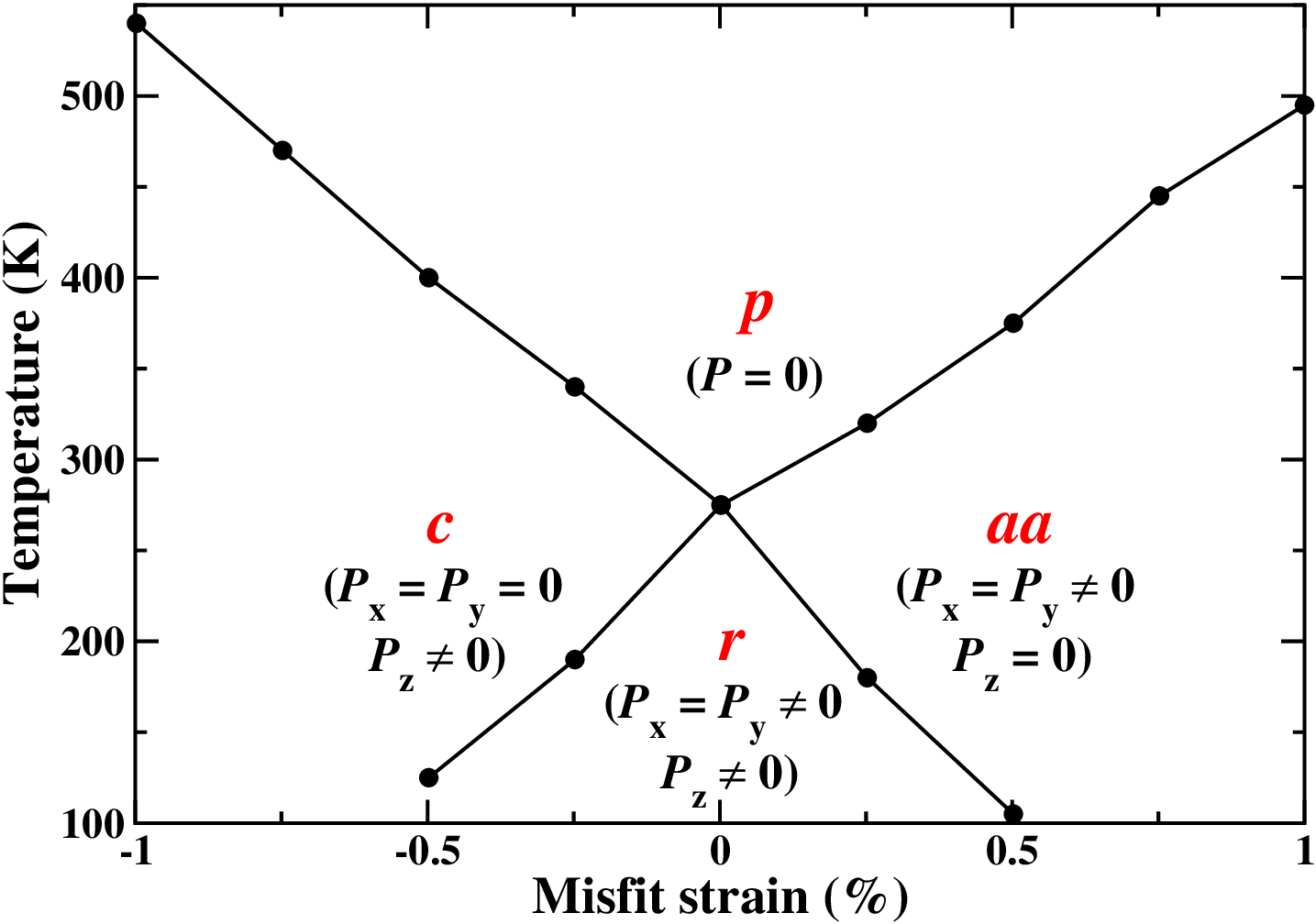}
\caption{Temperature-strain phase diagram for epitaxiallly strained
  BTO: transition temperatures are plotted as a function of misfit
  strain.  Here, $p$ denotes the high-temperature paraelectric phase,
  while $c$, $r$, and $aa$ denote the three ferroelectric phases with
  different polarization directions.}
\label{fig:temp-strain}
\end{figure}

To provide an appropriate reference for our calculations of the EC
effect in strained BTO, we first establish the strain-temperature
phase diagram for the Wu/Cohen-GGA parameterization of the effective
Hamiltonian. We calculate all components of the total polarization as
function of temperature for each value of strain, and from this we
identify the transition temperatures corresponding to the various
phase transitions. Similar to Ref.~\onlinecite{Dieguez_et_al_2004}, we
use the lattice constant immediately above the cubic to tetragonal
phase transition in the unstrained bulk calculation ($a_0 =
3.995$\,\AA) as reference to define zero misfit strain. The resulting
phase diagram is shown in Fig.~\ref{fig:temp-strain}.

The calculated phase diagram compares very well with the one obtained
in Ref.~\onlinecite{Dieguez_et_al_2004} using an LDA parameterization
of the effective Hamiltonian. Note that in contrast to
Ref.~\onlinecite{Dieguez_et_al_2004} we do not apply any empirical
pressure corrections in the effective Hamiltonian. At high
temperatures, strained BTO is paraelectric ($p$), while at lower
temperatures three different ferroelectric phases with different
orientations of the polarization vector $P$ occur ($c$, $r$ and $aa$,
see Fig.~\ref{fig:temp-strain}). All phase boundaries meet in one
point at $T \approx 270$\,K and zero strain, in good agreement with
previous
studies.~\cite{Pertsev_Zembilgotov_Tagantsev_1999,Dieguez_et_al_2004}

\begin{figure}[tb]
\centering
\includegraphics[width=\columnwidth]{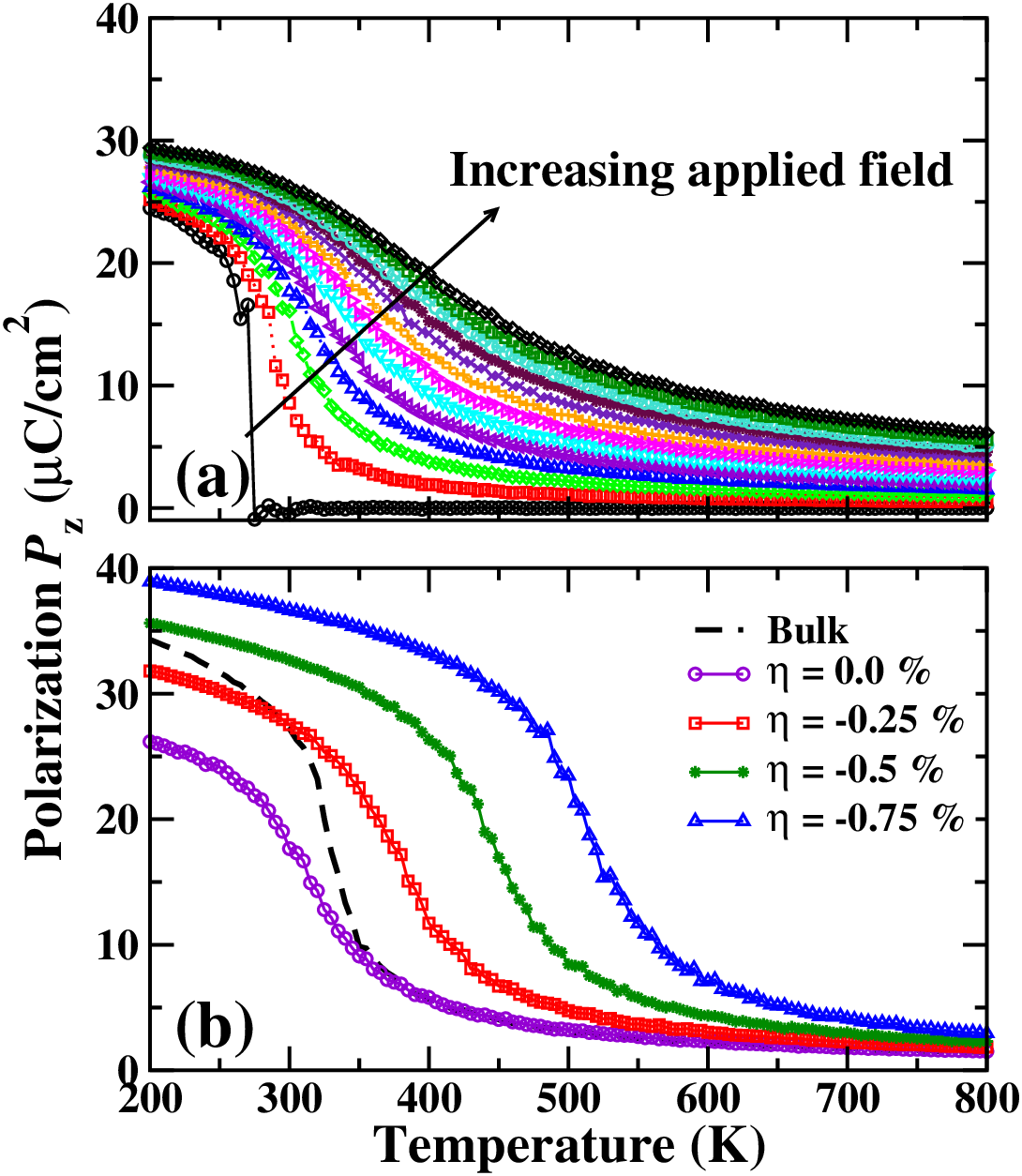}
\caption{(a) Polarization as function of temperature for strained
  BaTiO$_3$ under 0.0\,\% misfit strain at various applied fields
  starting from $0$ to $300$\,kV/cm in steps of 25\,kV/cm. (b)
  Polarization as function of temperature at an applied field of
  75\,kV/cm for unstrained bulk and clamped systems at different
  strains. Only the polarization component along the applied field
  direction, $P_z$, is plotted. }
\label{fig:Pz_vs_T-clamped}
\end{figure}

Next, we compute the EC temperature change $\Delta T$ at different
working temperatures for bulk BTO and mechanically clamped BTO under
different compressive strains (including zero strain). We use an
indirect approach to calculate the EC temperature change (see
e.g. Refs.~\onlinecite{Ponomareva/Lisenkov:2012} and
\onlinecite{Beckman_et_al_2012}). This approach is based on the
following Maxwell relation:
\begin{equation} 
\left.\frac{\partial S}{\partial E}\right|_T = \left.\frac{\partial P}{\partial
  T}\right|_E,
\end{equation}
which relates the isothermal entropy change due to an applied electric
field to the temperature dependence of the electric polarization at
constant electric field. The latter can be calculated using the
effective Hamiltonian approach. Here, $S$ and $T$ denote the entropy
and temperature of the system, $P$ is the total polarization, and $E$
is the applied electric field. The EC temperature change $\Delta T$ is
then given by:
\begin{equation} 
\label{eq:delta-T}
\Delta T = - \int_{E_1}^{E_2} \frac{T}{C_{p,E}} \Bigg(\frac{\partial
  P}{\partial T}\Bigg)_E \text{d}E,
\end{equation}
where $C_{p,E}$ is the specific heat at constant pressure and constant
electric field. We have used a temperature-independent value of $C_p =
2.53$\,Jcm$^{-3}$K$^{-1}$ for the specific heat, which corresponds to
the experimental value measured at room temperature.~\cite{He_2004}
$E_1$ and $E_2$ are the initial and final electric fields. We focus on
the temperature range of the paraelectric to ferroelectric transition,
i.e. $p$ to $r$ ($c$) for zero (nonzero) strain in
Fig.~\ref{fig:temp-strain}, where the strongest EC effect can be
expected.\cite{Ponomareva/Lisenkov:2012,Valant_2012,Rose/Cohen:2012}

Fig.~\ref{fig:Pz_vs_T-clamped}(a) shows the calculated temperature
dependence of the electric polarization for zero applied strain,
i.e. pure clamping, and different electric fields up to a maximum of
300\,kV/cm. Fields of this magnitude are typically applied in thin
film samples.\cite{Mischenko_et_al_2006,KarNarayan/Mathur:2010} The
external field is applied along the [001] direction,
i.e. perpendicular to the direction of clamping. There is a sharp
phase transition from paraelectric to ferroelectric ($r$-phase) in
absence of the external field. In presence of the field, the
transition temperature first shifts to slightly higher temperatures
and then the phase transition disappears and the system shows a smooth
dependence of polarization on temperature (compare also to
Refs.~\onlinecite{Rose/Cohen:2012} and
\onlinecite{Akcay_et_al_2007}). Qualitatively similar behavior is observed
for the free bulk case and for other strain values, only the
corresponding zero-field transition temperature is shifted according
to Fig.~\ref{fig:temp-strain} and the magnitude of the low temperature
polarization increases for increasing strain as shown in
Fig.~\ref{fig:Pz_vs_T-clamped}(b).

Note that the results in Fig.~\ref{fig:Pz_vs_T-clamped} are obtained
from statistical averaging over many MD steps and therefore contain
small statistical fluctuations. To compute $\partial P/\partial T$, we
therefore fit the polarization versus temperature using smoothing
cubic spline functions.  Due to the rather sharp transition for small
electric fields (first order in the free bulk case) the derivative is
ill-defined around the transition temperature, and we therefore use
$E_1 = 75$\,kV/cm and $E_2 = 300$\,kV/cm for the calculation of the EC
temperature change $\Delta T$ in Eq.~(\ref{eq:delta-T}), i.e. we
exclude small electric fields.

\begin{figure}[tb]
\centering \includegraphics[width=\columnwidth]{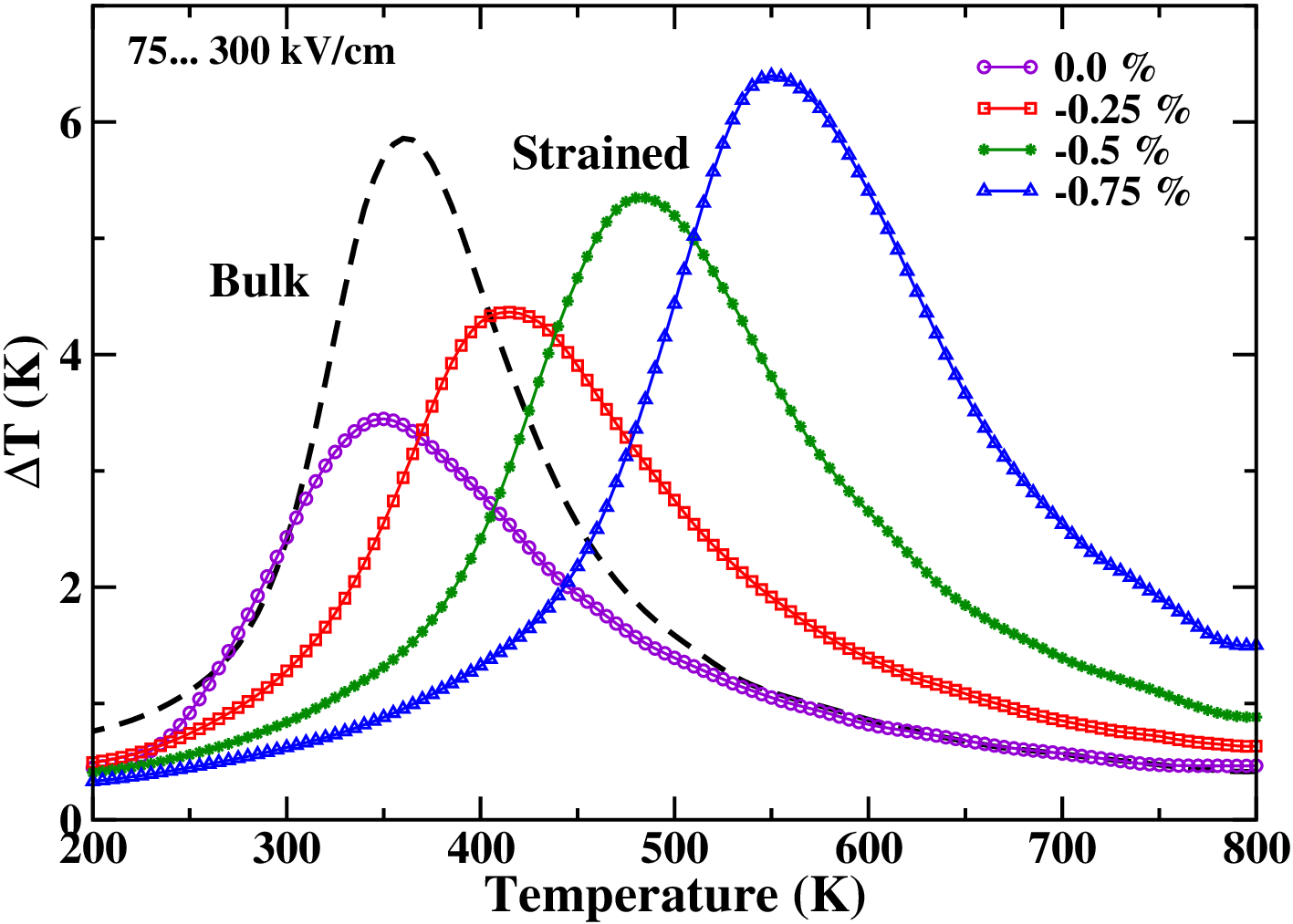}
\caption{The electrocaloric temperature change as a function of
  temperature: shown here for free (dashed line) and strained bulk
  BaTiO$_3$ with misfit strain equal to $0.0$\,\% (circles),
  $-0.25$\,\% (squares), $-0.5$\,\% (stars) and $-0.75$\,\%
  (triangles).  $\Delta T$ is calculated for a variation of the
  applied field from 75\,kV/cm to 300\,kV/cm. }
\label{fig:deltaT}
\end{figure}

The resulting $\Delta T$ is shown in Fig.~\ref{fig:deltaT} for the
various cases. For bulk BTO, a maximum $\Delta T$ of about $5.9$\,K is
reached at $T\approx 360$\,K, i.e. $65$\,K above the calculated zero
field transition temperature. This agrees well with previous
calculations.~\cite{Beckman_et_al_2012} Clamping by the substrate (at
zero strain) strongly reduces the EC temperature change to about
$3.4$\,K, consistent with previous results based on phenomenological
theory.~\cite{Akcay_et_al_2007} Note that for clamped BTO at zero
strain, the ferroelectric phase ($r$) has a non-zero polarization in
all three crystallographic directions. However, only the component
along the applied field direction ($P_z$) contributes to the EC
temperature change.

Increasing compressive misfit strain leads to a strong increase of the
peak value of $\Delta T$ compared to the clamped zero strain
case. Simultaneously, the temperature at which the maximum $\Delta T$
is observed also increases. This shift of the peak position of $\Delta
T$ is due to the increase of the paraelectric-to-ferroelectric
transition temperature with compressive strain (see
Fig.~\ref{fig:temp-strain}), since the largest $\Delta T$ is always
observed just above the ferroelectric transition temperature,
i.e. where the slope $\partial P / \partial T|_E$ is steepest.  The
increase of the maximum $\Delta T$ with compressive strain can also be
related to the increase of the ferroelectric transition temperature,
since it is to a large extent due to the factor $T$ in
Eq.~(\ref{eq:delta-T}). Physically, this means that, according to the
relation $\Delta Q = T \Delta S$, the amount of heat $\Delta Q$
connected with a certain entropy change $\Delta S$ is larger at higher
temperatures, and therefore the EC temperature change is larger at
higher temperatures (for the same $\Delta S$). In addition, there is a
small increase in $\partial P / \partial T|_E$ with compressive
strain, which is related to the increase of the spontaneous
polarization with compressive strain (see
Fig.~\ref{fig:Pz_vs_T-clamped}(b)), and which (to a smaller extent)
also contributes to the increase of $\Delta T$.

Interestingly, at strain values as small as $-0.75$\,\%, the maximum
$\Delta T$ becomes larger than in the bulk case, and reaches a value
of 6.4\,K. Previous experimental studies have reported that BTO films
with thicknesses up to 200\,nm can be grown coherently for epitaxial
strains as large as
$-$1\,\%,\cite{Choi_et_al_2004} which implies
that a large EC effect can be expected in epitaxially grown films of
BTO. We also note that, due to the strong shift of the transition
temperature, strain can also provide an effective tool to optimize the
EC effect at a given operating temperature. For example, for a device
that has to operate around 400\,K, strained BTO with $\eta =
-0.25$\,\% would be the best choice, even though a larger peak value
of $\Delta T$ can be achieved for $\eta = -0.75$\,\%.

One should note that while a large and tunable EC effect can be
achieved in thin films, a major obstacle for using thin films in
cooling devices is that the available volume of the EC material is
very small and therefore only small amounts of heat can be
transferred. One possibility to circumvent this problem could be the
use of multilayer geometries such as e.g. the one discussed in
Ref.~\onlinecite{KarNarayan/Mathur:2010}.

In summary, we have used MD simulations for a first principles-based
effective Hamiltonian to study the EC effect in epitaxially strained
bulk BTO. We have shown that clamping by the substrate (at zero
strain) strongly reduces the EC effect compared to bulk BTO, whereas
moderate compressive strain leads to a strong enhancement of the EC
effect as well as a strong shift of the temperature at which the
maximal EC temperature change is observed. Thus, misfit strain can be
utilized in two ways -- (i) to enhance the EC temperature change and
(ii) to achieve a maximal effect in a temperature range of interest
for a given application.

This work is supported by the Swiss National Science Foundation and
the German Science Foundation under the priority program SPP 1599
(``ferroic cooling''). We thank Takeshi Nishimatsu, Anna Gr\"unebohm,
and Peter Entel for useful discussions and help with the FeRAM code.

\bibliography{references}

\end{document}